\documentclass[paper]{JHEP}
\input epsf
\usepackage{cite}

\newcommand\NLO   {next-to-leading order }
\newcommand\as[1] {\ensuremath{\alpha_s^{#1}}}
\newcommand\asms  {\ensuremath{\alpha_s^{\overline{{\rm MS}}}}}
\newcommand\aeff  {\ensuremath{\alpha_{\rm eff}}}
\newcommand\aspt  {\ensuremath{\alpha_s^{\rm pert}}}
\newcommand\ep    {\ensuremath{\varepsilon}}

\newcommand\Fpow  {\ensuremath{F^{\rm pow}}}
\newcommand\Fpert {\ensuremath{F^{\rm pert}}}
\newcommand\muI   {\ensuremath{\mu_{\rm I}}}
\newcommand\epem  {\ensuremath{e^+e^-}}
\newcommand\Qb    {\ensuremath{{\bar Q}}}

\newcommand{\Nf}  {\ensuremath{{N_{\rm f}}}}
\newcommand\ms    {\ensuremath{\overline{{\rm MS}}}}

\newcommand\half  {\ensuremath{{\textstyle \frac12}}}
\newcommand\cf    {\ensuremath{{\cal F}}}
\newcommand\dcf   {\ensuremath{\delta{\cal F}}}
\newcommand\cm    {\ensuremath{{\cal M}}}
\renewcommand\O   {\mathrm O}

\renewcommand\to  {\ensuremath{\rightarrow}}
\renewcommand\d   {{\rm d}}
\renewcommand\i   {{\rm i}}

\newcommand{\beq} {\begin{equation}}
\newcommand{\eeq} {\end{equation}}
\newcommand{\beqa}{\begin{eqnarray}}
\newcommand{\eeqa}{\end{eqnarray}}
\newcommand\nn    {\nonumber}
\newcommand\bom[1]{{\mbox{\boldmath $#1$}}}

\title{Calculation of Power Corrections to Hadronic Event Shapes of
Tagged $\bom{b}$ Events}
\author{Zolt\'an Tr\'ocs\'anyi\thanks{Sz\'echenyi Fellow of the Hungarian
Ministry of Education. On leave from Department of Experimental
Physics, Kossuth University and Institute of Nuclear Research of the
Hungarian Academy of Sciences, Debrecen, Hungary} \\
Theory Division, CERN, CH-1211 Geneva 23, Switzerland}
\abstract{
We compute power corrections to mean values of hadronic event shapes ---
the thrust and the $C$ parameter --- of tagged $b$ quark events in
electron positron annihilation, using the dispersive approach. We find
that the leading power corrections are of the same type of $1/Q$
corrections as for event shapes in the massless case, with the same
non-perturbative coefficient times a perturbatively calculable
mass-dependent coefficient. The effect of the mass correction in the
power correction is to reduce the latter by 10--30\,\% for tagged $b$
events, for centre-of-mass energies ranging from the $Z^0$ peak down to
20\,GeV.}

\preprint{CERN--TH/99-340\\ hep-ph/9911353}
\begin{document}

\section{Introduction}

In recent years the idea of an infrared-regular effective strong coupling
at low scales (the strong coupling `freezing' \cite{DWevshape} or, more
rigorously, the `dispersive approach' of Dokshitzer, Marchesini and
Webber \cite{DMWdisp})  has been employed for estimating hadronization
corrections to various hadronic event shapes, using perturbative
calculations \cite{DWevshapedist,DLMSevshape,DLMSMilan,DMWeec,D-GWdis}.
These corrections arise in the form of power
corrections of order 1/$Q^p$, where $Q$ is the centre-of-mass energy.
Although the magnitude of the correction cannot be predicted by
perturbative means, the power $p$ and the relative coefficients among
different observables can be calculated using the assumption of
infrared freezing. This universality hypothesis has proved to give a
phenomenologically fairly consistent picture of power corrections
\cite{DELPHI,H1,JADE,Wickeqcd97,PAMFqcd98,SZqcd99}, and the same results
could be derived using a different method, the so-called renormalon approach
\cite{BBrenormalon,Brenormalon}.

In this letter we explore a further check of the universality picture.
Assuming the flavour independence of gluon radiation off quarks, the
analytic structure of the strong coupling in the infrared, defined in
an appropriate way, should not depend on the masses of the quarks, which
radiate the gluon. Therefore, in the calculation of the power
corrections to hadronic event shapes of tagged $b$ events, the mass of
the heavy quark enters in a completely perturbative manner. One may
expect that the leading mass correction could be $m_b/Q$ times the
leading power correction ($m_b$ being the heavy-quark mass) with some
calculable coefficient. If this coefficient is several times unity, then
the mass corrections in the power corrections are expected to be
significant enough to be a measurable effect.

A similar problem has already been considered by Nason and Webber in
calculating non-perturbative corrections to heavy-quark fragmentation in
\epem\ annihilation \cite{NWfragment}. Although the calculations are very
similar, there is an essential difference in the results: for the
fragmentation function there is a 1/$m_b$ power correction, while for
tagged events shapes there are mass corrections to the magnitude of the
leading 1/$Q^p$ power correction, which, however, vanish smoothly in the
massless limit.

Estimating the mass corrections in the power correction may be interesting
from another point of view, too. Recently, \NLO calculations of
three-jet quantities in electron positron annihilation have been
performed in which quark mass effects have been taken into account
explicitly \cite{BBU3jet,RSB3jet,NO3jet}. Using this theoretical input
and the value of the $b$-quark mass, the flavour independence of the
strong interactions was demonstrated by determining the ratio of strong
couplings, \as{b}/\as{uds} \cite{SLDflindep, OPALflindep}. One can turn
around the argument and, assuming flavour independence, the $b$-quark mass
can be measured from the sensitivity of three-jet event shapes to mass
effects \cite{DELPHIbmass,BBMOUbmass,ALEPHbmass}. Such a measurement requires
an estimate of hadronization corrections. The traditional way of obtaining
such estimates is by Monte Carlo event generators \cite{JETSET,HERWIG}.
Those programs use the heavy-quark mass as input. As a result, the
estimate of hadronization corrections brings a significant systematic
error into the measurement of $m_b$. For taking into account the
hadronization correction in a different way, the simple formulae of power
corrections presented in this paper can easily be incorporated in a fit
to the $b$-quark mass, which would hopefully result in a more stringent
mass measurement.

As mentioned above, the presence of the quark mass appears only in the
perturbative calculation. Therefore, the calculations that lead to the
appearance of the Milan factor in the power corrections \cite{DLMSMilan}
can also be performed in the presence of the quark mass, resulting in a
mass-dependent Milan factor. In this paper we make only the first step
and calculate that part of power corrections to the mean value of two
\epem\ event shapes, the thrust and the $C$ parameter, which was termed
`naive contribution' in Ref.~\cite{DLMSMilan}.

\section{The dispersive approach}

In employing the dispersive approach of Ref.~\cite{DMWdisp} one starts
with assuming the validity of the following dispersion relation for the
strong coupling:
\beq\label{spectral}
\as{}(k^2) = -\int_0^\infty\!\frac{\d\mu^2}{\mu^2+k^2}\,\rho_s(\mu^2)\:,
\qquad
\rho_s(\mu^2) = -\frac{1}{2\pi\i}{\rm Disc}\left\{\as{}(-\mu^2)\right\}\:.
\eeq
Further, it is also assumed that similarly to QED, the dominant
effect of the running of \as{}\ on some QCD observable $F$ may be
represented in terms of the spectral function $\rho_s(\mu^2)$ and a
characteristic function $\cf(\mu^2)$:
\beqa
&&
F = \as{}(0)\,\cf(0;\ldots) +
\int_0^\infty\!\frac{\d\mu^2}{\mu^2}\,\rho_s(\mu^2)\,\cf(\mu^2;\ldots)
\\ && \quad \label{F}
= \int_0^\infty\!\frac{\d\mu^2}{\mu^2}\,\rho_s(\mu^2)\,
\left[\cf(\mu^2;\ldots) - \cf(0;\ldots)\right]\:,
\eeqa
where we used Eq.~(\ref{spectral}) to express \as{}(0). The
characteristic funtion $\cf(\mu^2;\ldots)$ is a function of dimensionless
ratios of $\ep\equiv \mu^2/Q^2$, $\rho \equiv 4\,m_b^2/Q^2$, where $m_b$
is the heavy-quark mass and $\{y\}$, the collection of any further
relevant dimensionless parameters, e.g.\ the jet shape $S$ in our
present considerations. We obtain \cf\ by computing the one-loop
graphs, corresponding to the physical process under consideration, with
a non-zero gluon mass $\ep\ne 0$, and dividing by \as{}:
\beq\label{calF}
\cf(\ep; \rho, S) = 
\int\!\d\Phi(\{x_i\}; \rho, \ep)\,\cm(\{x_i\}; \rho, \ep)\,
S(\{x_i\}; \rho, \ep)\:,
\eeq
where $\d\Phi(\{x_i\}; \rho, \ep)$ denotes the phase space in terms of
the independent variables $\{x_i\}$, $\cm$ is the proper squared
amplitude divided by \as{}, and $S$ stands for the event shape variable.
Introducing the effective coupling $\aeff(\mu^2)$, defined in terms of
the spectral function by
\beq\label{aeff}
\rho_s(\mu^2) = \mu^2 \frac{\d\aeff}{\d\mu^2}\:,
\eeq
we can integrate Eq.~(\ref{F}) by parts to obtain
\beq
F =
\int_0^\infty\!\frac{\d\mu^2}{\mu^2}\,\aeff(\mu^2)\,\dot{\cf}(\mu^2) =
\int_0^\infty\!\frac{\d\ep}{\ep}\,\aeff(Q^2 \ep)\,\dot{\cf}(\ep)\:,
\eeq
where
\beq\label{cfdot}
\dot{\cf}(\ep) = - \ep \frac{\d\cf}{\d\ep}\:.
\eeq
Using the definition Eq.~(\ref{aeff}) and the dispersion relation
Eq.~(\ref{spectral}), we can deduce that in the perturbative domain,
$\as{} \ll 1$, the standard and effective couplings are approximately the
same \cite{DMWdisp}:
\beq\label{asequality}
\aeff(\mu^2) = \as{}(\mu^2) + \O(\as{3})\:.
\eeq
Thus we may interpret \aeff\ as an effective coupling that extends the
physical perturbative coupling into the non-perturbative domain. If the
effective coupling has a non-perturbative component
$\delta\aeff(\mu^2)$, with support limited to low values of $\mu^2$,
the corresponding contribution to $F$,
\beq
\delta F = \int_0^\infty\!\frac{\d\ep}{\ep}\,
\delta\aeff(Q^2 \ep)\,\dot{\cf}(\ep)\:,
\eeq
will have a $Q^2$ dependence determined by the small-$\ep$ behaviour of
$\dot{\cf}$.

We shall be interested in the leading power behaviour of $\delta F$,
which will be called the leading power correction to $F$,
\beq\label{Fpow}
\delta F = \Fpow + O\Big(1/Q^p\Big)\:.
\eeq
In order to determine \Fpow, we recall that power-suppressed
contributions to $\delta F$ can arise from only those terms in \cf\
that are non-analytic at $\ep = 0$ \cite{DMWdisp}. Therefore, the
leading power correction \Fpow\ will be obtained from the leading
non-analytic term at $\ep = 0$ in $\dot{\cf}$, which we denote by
$C_F\,\dcf/2\pi$.

\section{Calculations}

In calculating the function \dcf\ for the \epem \to $\:Q\Qb g$ event
shapes, we find three sources of quark-mass dependence:
(i) restriction of phase space;
(ii) mass corrections in the matrix element;
(iii) mass corrections in the definition of the event shape.
The double differential cross section for the production of a massive
quark antiquark pair and a massive gluon given in terms of the scaled
quark, antiquark and gluon energies, $x$, $\bar{x}$ and
$x_g = 2 - x - \bar{x}$ was  derived in Ref.~\cite{NWfragment}. This
cross section is an analytic function of the gluon mass and, therefore, in
calculating \dcf, the \ep-dependent terms can be dropped completely. 
In the case of the vector current contribution the cross section for zero
gluon mass is given by 
\beqa
\label{sigmabbgV}
&&
\frac{1}{\sigma_V}\frac{\d \sigma_V}{\d x\,\d \bar{x}} =
\frac{C_F}{\beta}\frac{\as{}}{2\pi}
\Bigg[
\frac{-\lambda(x^2 - \rho, \bar{x}^2 - \rho, x_g^2)}
     {8\,(1 - x)^2\,(1 - \bar{x})^2}
\\ \nn && \qquad\qquad\qquad\qquad
 + \frac1{1+\half\rho}\,
\left(\frac{1}{(1 - x)^2} + \frac{1}{(1 - \bar{x})^2}\right)
\Bigg]\:;
\eeqa
for the axial vector current contribution we have
\beqa
\label{sigmabbgA}
&&
\frac{1}{\sigma_A}\frac{\d \sigma_A}{\d x\,\d \bar{x}} =
\frac{C_F}{\beta}\frac{\as{}}{2\pi}
\Bigg[
\frac{-\lambda(x^2 - \rho, \bar{x}^2 - \rho, x_g^2)}
     {8\,(1 - x)^2\,(1 - \bar{x})^2}
\\ \nn && \qquad\qquad\qquad\qquad
 + \frac1{1-\rho}\,
\left(\frac{1}{(1 - x)^2} + \frac{1}{(1 - \bar{x})^2}
    + \frac\rho2\,\frac{x_g^2}{(1 - x)\,(1 - \bar{x})}\right)
\Bigg]\:,
\eeqa
where $\beta = \sqrt{1 - \rho}$ is the quark velocitey and
$\lambda(x, y, z) = x^2 + y^2 + z^2 - 2\,x\,y - 2\,y\,z - 2\,z\,x$.
In Eqs.~(\ref{sigmabbgV}) and (\ref{sigmabbgA})
\beq
\sigma_V = \sigma_0\,\left(1+\half\,\rho\right)\,\beta\:, \qquad
\eeq
and
\beq
\sigma_A = \sigma_0\,\left(1-\rho\right)\,\beta
\eeq
are the Born cross sections for heavy-quark production by a vector and an
axial vector current respectively, $\sigma_0$ being the massless quark
Born cross section. The corresponding phase space was also given in
Ref.~\cite{NWfragment}:
\beq
\label{psbbg}
\d\Phi(x, \bar{x}; \ep, \rho) = \d x\,\d\bar{x}\,
\Theta\Big(-{\textstyle \frac18}\,
\lambda(x^2 - \rho, \bar{x}^2 - \rho, x_g^2 - 4 \ep)\Big)\,
\eeq
The phase-space boundary in this equation gives $x_- \le \bar{x} \le
x_+$, where
\beq
x_\pm = \frac{(2 - x)(1 - x- \ep + \half\,\rho) \pm
\sqrt{(x^2 - \rho)[(1 - x - e)^2 - \ep\,\rho]}}{2\,(1 - x) + \half\,\rho}
\eeq
and
\beq
\sqrt{\rho} \le x \le 1 - e - \sqrt{\ep\,\rho}\:.
\eeq
For large values of the gluon momentum the phase-space boundary does
not yield leading non-analytic contributions in $\ep$.  Therefore, the
leading non-analytic contribution to \dcf\ comes from the soft-gluon
emission region \cite{DMWdisp,NWfragment}, where both the matrix
elements, in Eqs.~(\ref{sigmabbgV}) and (\ref{sigmabbgA}), and the phase
space, in Eq.~(\ref{psbbg}), can be expanded in $x_g = y + \bar{y} \simeq
\sqrt{\ep}$. This expansion for the cross section formulae results in
\beq\label{sigmaexpanded}
\frac{1}{\sigma_{V,A}}\frac{\d \sigma_{V,A}}{\d y\,\d \bar{y}} \approx
\frac{C_F}{\beta}\frac{\as{}}{2\pi}
\left[\frac{2}{y\,\bar{y}}
- \frac\rho2\,\frac{(y+\bar{y})^2}{y^2\,\bar{y}^2}\right]
\eeq
both for vector and axial vector current contributions, where
$y = 1 - \bar{x}$ and  $\bar{y} = 1 - x$.  The lower boundary of the
phase space in $x_g = y + \bar{y} \simeq \sqrt{\ep}$ region is
approximated  as follows:
\beq
-\frac18\,\lambda(x^2 - \rho, \bar{x}^2 - \rho, x_g^2 - 4 \ep)
\approx
2\,y\,\bar{y} - \frac\rho2\,(y+\bar{y})^2 - 2\,\ep\,(1- \rho) = 0\:.
\eeq
Thus, to obtain \dcf\ for mean values of event shapes, we have to
calculate the following integral
\beqa\label{dcF}
&&\!\!\!\!\!\!
\dcf =
-\,\frac\ep\beta\,\frac{\d}{\d\ep}
\int\!\d y\,\d\bar{y}\,
\Theta(y - \sqrt{\ep\rho})\,\Theta(\bar{y} - \sqrt{\ep\rho})\,
\Theta(1 - \rho/2 - y - \bar{y})
\\ \nn && \qquad\qquad\qquad\;
\times
\Theta(4\,y\,\bar{y} - \rho\,(y+\bar{y})^2 - 4\,\ep\,(1- \rho))
\\ \nn && \qquad\qquad\qquad\;
\times
\frac{2}{y\,\bar{y}}\,\left(
1 - \frac\rho4\,\frac{(y+\bar{y})^2}{y\,\bar{y}}\right)
S(y, \bar{y}; \rho, \ep)\:.
\eeqa
In this equation, the exact form of the upper boundary does not
influence those non-analytic terms that give the leading power
correction.

The mass corrections in the physical quantity introduce an
observable-dependent, but perturbatively calculable mass dependence.
As simple application of formula (\ref{dcF}), we consider two event shapes,
the $C$ parameter and the thrust $T$.  The $C$ parameter \cite{Cpar} is
derived from the eigenvalues of the infrared-safe momentum tensor
\beq
\theta^{ij} =
\sum_a \frac{p_a^i p_a^j}{|\vec{p}_a|}\bigg/\sum_a |\vec{p}_a|,
\eeq
where the sum on $a$ runs over all final-state hadrons and $p_a^i$ is the
$i$th component of the three-momentum $\vec{p}_a$ of hadron $a$ in the
c.m.\ system. The tensor $\theta$ is normalized to have unit trace. In
terms of the eigenvalues $\lambda_i$ of the $3 \times 3$ matrix
$\theta$, the global shape parameter $C$ is defined as
\beq\label{Cdef}
C =
3\,(\lambda_1 \lambda_2 + \lambda_2 \lambda_3 + \lambda_3 \lambda_1)\:.
\eeq
The thrust \cite{Thrust} is given by
\beq
\label{thrust}
T=\max_{\vec{n}_T}
\left(\frac{\sum_a |\vec{p}_a\cdot\vec{n}_T|}{\sum_a |\vec{p}_a|}\right),
\eeq
where the sum runs over all final-state particles and the thrust axis
${\vec{n}_T}$ is chosen to maximize the expression. For three partons in
the final state, the thrust can be written as \cite{CTTWresum}
\beq
T = 2
\frac{\max\{|\vec{p}_1|, |\vec{p}_2|, |\vec{p}_3|\}}{\sum_a |\vec{p}_a|}\:.
\eeq
Instead of the thrust, we shall consider its deviation from unity, $t = 1-T$.

For these two event shapes the contribution of the non-perturbative
gluon will just add to that of the underlying perturbative event
(contribution of many soft perturbative gluons), and the power
correction based on the presence of just a single non-perturbative
gluon will remain valid \cite{SZqcd99}. This argument can be used
independently of the mass of the leading quark pair. As discussed in
Ref.~\cite{DLMSMilan}, the `naive contribution' to the mean value of the
$C$ parameter is obtained from Eq.~(\ref{Cdef}) in the soft-gluon
approximation, with gluon mass set to zero. The corresponding formula
with non-zero quark masses is the following:
\beq\label{Capprox}
C(y, \bar{y}; \rho) \approx
\frac{6}{\beta^3}\,\frac{y\,\bar{y}}{y+\bar{y}}\,
\left(1 - \frac\rho4\,\frac{(y+\bar{y})^2}{y\,\bar{y}}\right)\:.
\eeq
Similarly, we need only the leading term in the soft-gluon expansion
of the function $t(y, \bar{y}; \rho)$ with the gluon mass neglected,
\beq\label{tapprox}
t(y, \bar{y}; \rho) \approx \frac12\,\left(
  \frac{y+\bar{y}}{\sqrt{1-\rho}}
- \frac{1}{1-\rho}\,
  \Big[(\bar{y} - y)\,\Theta(\bar{y} - y)
     + (y - \bar{y})\,\Theta(y - \bar{y})\Big] \right)\:.
\eeq

To perform the integrations in Eq.~(\ref{dcF}) we introduce polar
coordinates, $y = r\,\cos \phi$, $\bar{y} = r\,\sin\phi$. In terms of the
variables $r$ and $\phi$
\beq
\d\Phi(\ep) = r\,\d r\,\d\phi\,
\Theta(\phi-\delta)\,\Theta(\half\pi-\delta-\phi)\,
\Theta(r_+(\phi) - r)\,\Theta(r - r_-(\phi,\ep))\:,
\eeq
where 
\beq
r_+(\phi) = \frac{(1 + \sin 2\delta)^{-1}}{\sqrt{1 + \sin 2\phi}}\:,
\quad
r_-(\phi,\ep) =
\sqrt{\frac{2\,\ep\,(1 - \sin 2\delta)}{\sin 2\phi - \sin 2\delta}}\:,
\quad
\sin 2\delta = \frac{\rho}{2-\rho}\:.
\eeq
The $r$ integral is trivial and the integral over $\phi$ can be performed
after making the shift $\phi\to\phi-\pi/4$.  After performing the
differentiation in Eq.~(\ref{dcF}), for the $C$ parameter,
$S = C$, we obtain
\beqa\label{dcfC}
&&
\dcf^{(C)} = 
6\pi\,\sqrt{\ep}\,\frac1\beta\,\left(
\frac{1}{1+\sqrt{\rho}} - \frac{\sqrt{\rho}}{2}\right)
\\ \label{dcfCexpanded} && \qquad\quad
= 6\pi\,\sqrt{\ep}\,\left(1-\frac32\,\sqrt{\rho} + \O(\rho)\right)\:.
\eeqa
From the expansion at small values of $\rho$, we see that our formula
reproduces the known zero mass result \cite{DWevshape}.
In the case of thrust, $S = t$, one finds
\beq\label{dcft}
\dcf^{(t)} = 
\sqrt{\frac\ep\rho}\left(\pi\,\beta 
- \frac{1}{1-\rho}\Big(
\alpha_+ + \alpha_- - 2\,\sqrt{\rho\,(1-\rho)}\Big)\right)\:,
\eeq
where 
\beq
\alpha_\pm = \arctan\frac{2 - \rho \pm \sqrt{2}}{\sqrt{\rho\,(1-\rho)}}\:.
\eeq
Expanding $\dcf^{(t)}$ in $\rho$ we find
\beq\label{dcftexpanded}
\dcf^{(t)} =
4\,\sqrt{\ep}\,\left(1 - \frac{3\pi}{8}\sqrt{\rho} + \O(\rho)\right)\:,
\eeq
which shows that the apparent $\sqrt{\ep/\rho}$ behaviour in
Eq.~(\ref{dcft}) is in fact a $\sqrt{\ep}$ behaviour with
multiplicative mass corrections that are regular for vanishing quark mass.
Setting $\rho = 0$, we find agreement with the known zero-mass result
\cite{DWevshape}. 

In Fig.~\ref{f:dcF} we plotted the functions $\dot{\cf}(\ep)$ in
Eq.~(\ref{cfdot}), obtained from numerical integration for representative
values of $\rho$, with \cf\ as given in Eq.~(\ref{calF}), but with $S$
taken in the soft-gluon approximation (Eq.~(\ref{Capprox}) for $S = C$ in
Fig.~\ref{f:dcF}a and Eq.~(\ref{tapprox}) for $S = t$ in
Fig.~\ref{f:dcF}b). In the same figure, the solid lines show the
analytic results $\dcf^{(C)}$, Eq.~(\ref{dcfC}), and $\dcf^{(t)}$,
Eq.~(\ref{dcft}). We clearly see the leading $\sqrt{\ep}$ behaviour for small
values of $\ep$ together with the correct quark-mass dependence.
\FIGURE{
\label{f:dcF}
\begin{minipage}[t]{.47\linewidth}
\centerline{ \epsfxsize=7.2cm \epsfbox{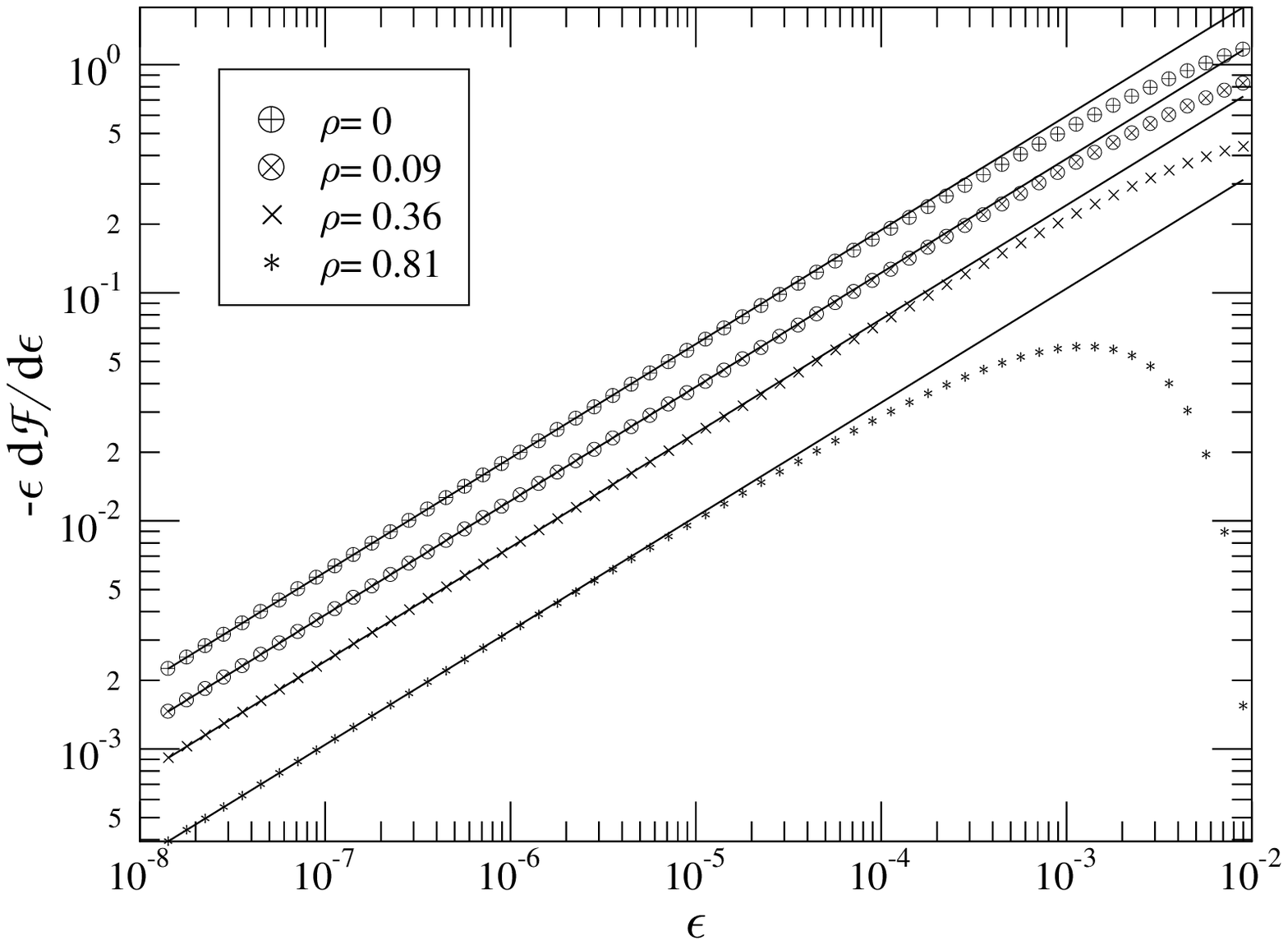} }
\end{minipage}
\hfill
\begin{minipage}[t]{.47\linewidth}
\centerline{ \epsfxsize=7.2cm \epsfbox{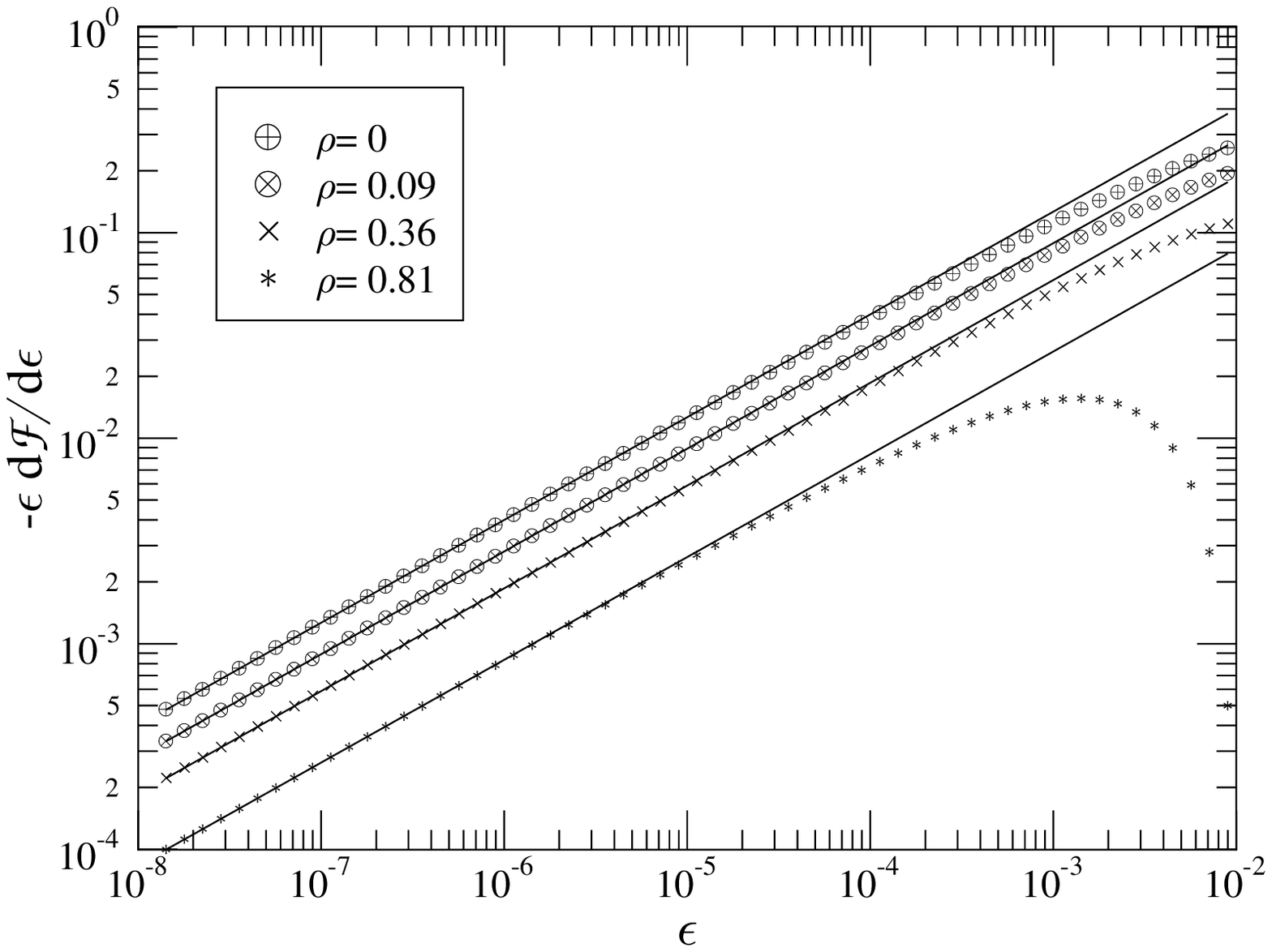} }
\end{minipage}
\caption{Derivative of the characteristic function for the mean value of
the (a) $C$ parameter (b) $1 - T$ for values of $\rho = 0$, 0.09, 0.36,
0.81.}
}

Having calculated the derivative of the characteristic functions for the
various event shapes, we can use Eq.~(\ref{Fpow}) to obtain the power
correction \Fpow. Following Ref.~\cite{DMWdisp}, we introduce the moment
integral
\beq
A_{2p} =
\frac{C_F}{2\pi}\int_0^\infty\!\frac{\d\ep}{\ep}\,\ep^p\,
\delta\aeff(\ep Q^2)\:.
\eeq
In terms of this non-perturbative parameter, the `naive contribution'
to the power corrections to the mean value of $C$ and $t$ are given by
the $p=\half$ moment of $\delta\aeff$ as
\beq\label{CpowLM}
\langle C\rangle^{\rm pow} =
6\,\pi\,\frac{A_1}{Q}\,\left(1 - 3\,\frac{m_b}{Q}
+ \O\left(\frac{m_b^2}{Q^2}\right)\right)
\eeq
and
\beq\label{tpowLM}
\langle t\rangle^{\rm pow} =
4\,\frac{A_1}{Q}\,\left(1 - \frac{3\pi}{4}\,\frac{m_b}{Q}
+ \O\left(\frac{m_b^2}{Q^2}\right)\right)\:,
\eeq
with the {\em same} value of the moment integral as phenomenologically
deduced from untagged samples.  To obtain these leading
mass corrections, valid for small values of the tagged-quark mass, we
used Eqs.~(\ref{dcfCexpanded}) and (\ref{dcftexpanded}). These results
show that at LEPI energies the mass correction reduces the magnitude of
the power correction by about 10\,\% for the mean value of the $C$
parameter and by about 7\,\% for the mean value of $t$ (with $m_b$ taken to
be the \ms\ mass at the centre-of-mass energy). 

A two-loop analysis of power corrections to event shapes in the
massless-quark case revealed that there is some freedom in the definition
of the `naive contribution'. With a different definition for this term,
the other two (`inclusive' and `non-inclusive') contributions change in
such a way that the sum of the three terms remains unambiguous. The universal
result can be summarized by a simple multiplicative correction factor,
the so-called Milan factor \cite{DLMSMilan}.\footnote{Recently , it was
pointed out that the correct value of the Milan factor is $\cm\approx
1.5$ \cite{D-GMSmilan}.} A two-loop calculation with mass effects taken
into account would modify our results in Eqs.~(\ref{CpowLM}) and
(\ref{tpowLM}) in two ways.  On the one hand, these equations would
acquire an overall Milan factor of 1.5 and on the other the
coefficients in front of the $m_b/Q$ corrections would change. For the
rest of this paper, we shall neglect the latter modification, but shall
include the Milan factor.

\section{Merging perturbative and non-perturbative contributions}

Once we have the leading power correction, we have to combine it with the
perturbative prediction for the same physical quantity to obtain the full
theoretical prediction:
\beq
F(Q) = \Fpert(Q) + \Fpow(Q)\:.
\eeq
The two contributions are separately ill-defined because the perturbative
part is given by an expansion that is factorially divergent, while in the
power correction, the non-perturbative part of the effective coupling
$\delta\aeff$ contains an ill-defined all-order subtraction of the
pure perturbative part off the full effective coupling \aeff. The sum of
the two contributions is finite. At fixed order in perturbation theory
they should be merged in such a way that the terms that would grow
factorially in an all-order result cancel order by order.  We follow
the prescription of Ref.~\cite{DLMSMilan} for this merging, where the
moment integral $A_1$ was approximated with the same integral up to an
infrared scale \muI,
\beq\label{A1approx}
A_1 \simeq \frac{2\,C_F}{\pi^2}
\int_0^{\muI}\!\d k\,\left[\as{}(k) - \aspt(k)\right]\:.
\eeq
In this equation the integrand is the non-perturbative component of the
strong coupling, i.e. the difference of the strong coupling \as{}\ as
given by Eq.~(\ref{spectral}) and the perturbative coupling \aspt. Above
the infrared scale \muI, this non-perturbative coupling is assumed to give
negligible contribution to the moment integral (see
Eq.~(\ref{asequality}) and the following paragraph).

The integral of the strong coupling in Eq.~(\ref{A1approx}) can be
expressed in terms of a phenomenological parameter $\alpha_0$,
\beq
\int_0^{\muI}\!\d k\,\as{}(k) \equiv \muI\,\alpha_0(\muI)\:.
\eeq
The value of this parameter depends on the infrared matching scale. 
For $\muI = 2\,$GeV, fits of event shapes of untagged events gave $\alpha_0
\simeq 0.5$ \cite{DWevshape,H1,Wickeqcd97,PAMFqcd98,SZqcd99}.

To calculate the integral of the perturbative coupling \aspt, we use its
one-loop expression  given as a geometric series:
\beq\label{aspt}
\frac{\aspt(k)}{2\pi} = 
\frac{\aspt(Q)}{2\pi} \sum_{\ell = 0}^\infty
\left(\frac{\aspt(Q)}{2\pi}\,\beta_0^{(4)}\,\ln\frac{Q}{k}\right)^\ell\:,
\eeq
where $\beta_0^{(4)}$ is the one-loop beta function
\beq
\beta_0^{(\Nf)} = \frac{11}{3}\,C_A - \frac23\,\Nf\:,
\eeq
for $\Nf = 4$ light fermion flavours, and $\aspt(Q)$ is a four-flavour
perturbative coupling in the physical (CMW) renormalization scheme
\cite{CMWscheme}.

For the practically interesting cases the perturbative prediction
for the event shape the perturbative prediction \Fpert\ is known to
second order in \as{},
\beq\label{Fpert}
\Fpert(Q; 2) = \frac{\asms(Q)}{2\,\pi}\,B(m_b, Q)
+ \left(\frac{\asms(Q)}{2\,\pi}\right)^2\,C(m_b, Q)
+ \O\left((\asms(Q))^3\right)\:,
\eeq
with Born and correction coefficients, $B(m_b, Q)$ and $C(m_b, Q)$
respectively, calculated in the \ms\ scheme at scale $Q$, and $\asms(Q)$
is the four-flavour strong coupling defined in the \ms\ renormalization
scheme.  In such cases the summation in Eq.~(\ref{aspt}) should be
truncated at O(\as{2}), that is at $\ell = 1$. Then the integral of the
perturbative coupling in Eq.~(\ref{A1approx}) gives
\beq\label{Int-aspt}
\muI\,\left[\frac{\aspt(Q)}{2\,\pi}
+\left(\frac{\aspt(Q)}{2\,\pi}\right)^2
\beta_0^{(4)}\,\left(\ln\frac{Q}{\muI} + 1\right)
\right]\:.
\eeq
In order to use the \ms\ coupling everywhere in the final result, one
has to make the shift
\beq
\aspt \to \asms\,\left(1 + K\,\frac{\asms}{2\,\pi}\right)
\eeq
in Eq.~(\ref{Int-aspt}), with $K$ defined as
\beq
K = C_A\left(\frac{67}{18}-\frac{\pi^2}{6}\right)-\frac59\,\Nf\:,\quad
\Nf = 4\:.
\eeq
The coupling in the definition of the 
non-perturbative parameter $\alpha_0$ is the physical coupling;
therefore, the value of $\alpha_0$ does not depend on the chosen scheme. 

Collecting the various contributions, for the event shape $S( = t$, or
$C)$ we find
\beqa\label{Fexpansion}
&&
F^{(S)}(Q) = \bar{\alpha}_s\, B + \bar{\alpha}_s^2\, C
\\ \nn && \qquad\qquad
+\frac{2}{\pi}\cm\,
\frac{2C_F}{\sqrt{\ep}}\,
\delta{\cal F}^{(S)}(\rho, \ep)\,
\frac{\muI}{Q}
\left(\bar{\alpha}_0 - \bar{\alpha}_s - \bar{\alpha}_s^2
\left[\beta_0^{(4)}\,\left(\ln\frac{Q}{\muI} + 1\right) + K\right]\right) 
\:,
\eeqa
where $\bar{\alpha}_s \equiv \asms(Q)/2\pi$ for $\Nf = 4$ flavours and
$\bar{\alpha}_0 \equiv \alpha_0/2\pi$. In order to use the standard $\Nf
= 5$ flavour \ms\ strong coupling, we have to make the shift
\beq
\bar{\alpha}_s \to 
\bar{\alpha}_s + \frac23\,\bar{\alpha}_s^2\,\ln\left(\frac{m_b}{Q}\right)\:.
\eeq
We could make a more sophisticated description of the heavy-quark threshold
in the running coupling (see Refs.~\cite{PDG98,DSthreshold}),
but the usual centre-of-mass energy for event shapes being far from the
threshold, the difference is negligible. For instance, using the
prescription of Ref.~\cite{DSthreshold} changes the physical prediction
$F^{(S)}(Q)$ by less than 1\,\% at $Q = 20\,$GeV and by about
$0.1\,\%$ at the $Z^0$ peak.  The renormalization scale
dependence of this result can be studied in the usual way. To show the
effect of the mass correction, we plot $F^{(S)}$ for $S = C$ in
Fig.~\ref{f:Fexpansion}a and for $S = t$ in Fig.~\ref{f:Fexpansion}b
for $\muI = 2\,$GeV, $\alpha_0 = 0.5$ and the world average of the strong
coupling at the $Z^0$ peak, $\as{}(M_Z) = 0.119$ \cite{Bethke}. The solid
lines show the results for the central value of the \ms\ $b$-quark mass
at the given hard scattering scale run from $m_b(m_b) = 4.3\,$GeV
\cite{PDG98}. The dotted line is the next-to-leading order
perturbative prediction, with mass effects included, and the dashed lines
represent the result when mass effects are present in the
perturbative prediction plus power corrections without mass effects. 
The perturbative coefficients $B$ and $C$ were obtained using the {\tt
zbb4} program \cite{zbb4}.  We can observe that the mass effect in the
power correction in tagged $b$ samples is important for centre-of-mass
energies below about 45\,GeV. 
\FIGURE{
\label{f:Fexpansion}
\begin{minipage}[t]{.47\linewidth}
\centerline{ \epsfxsize=7.2cm \epsfbox{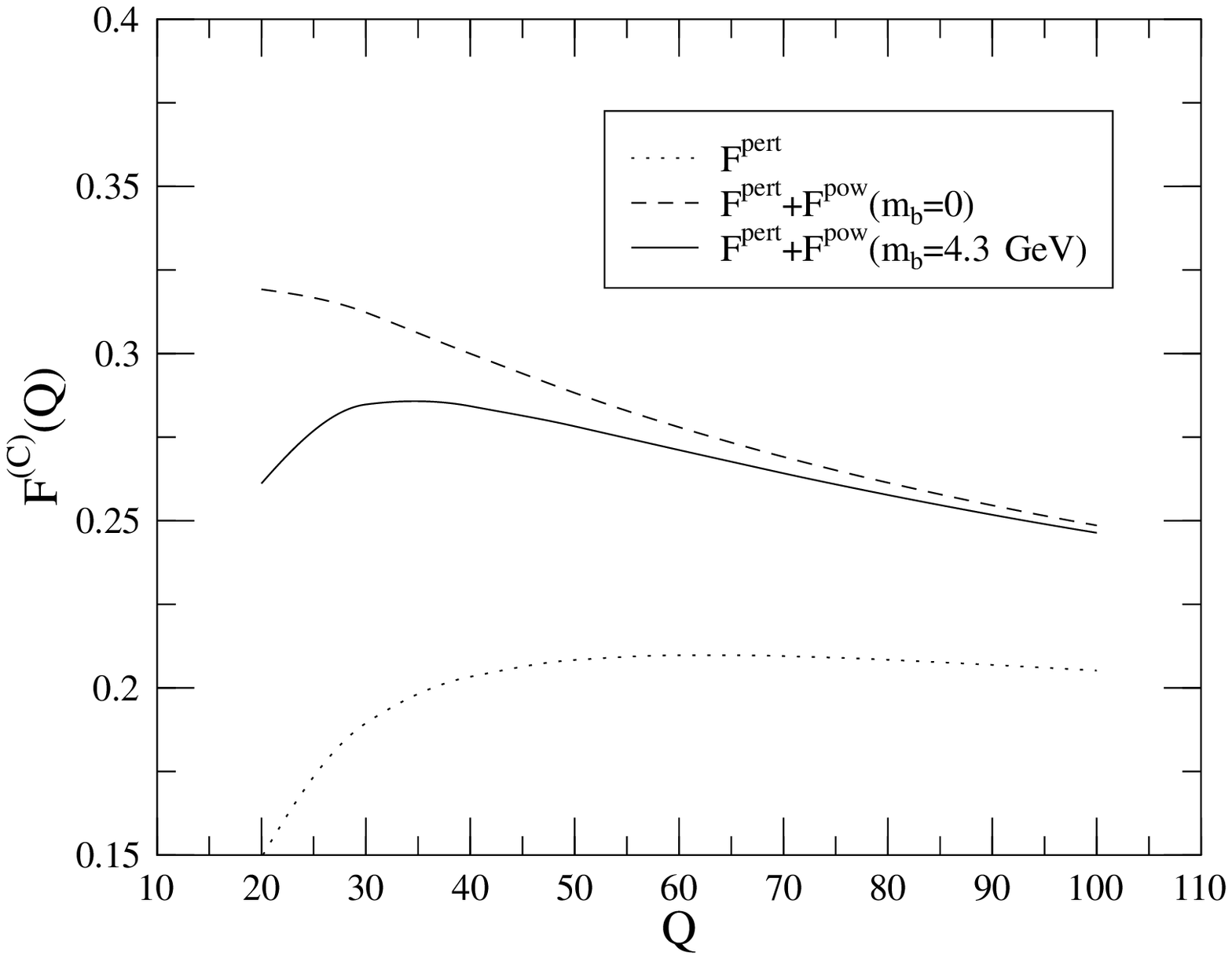} }
\end{minipage}
\hfill
\begin{minipage}[t]{.47\linewidth}
\centerline{ \epsfxsize=7.2cm \epsfbox{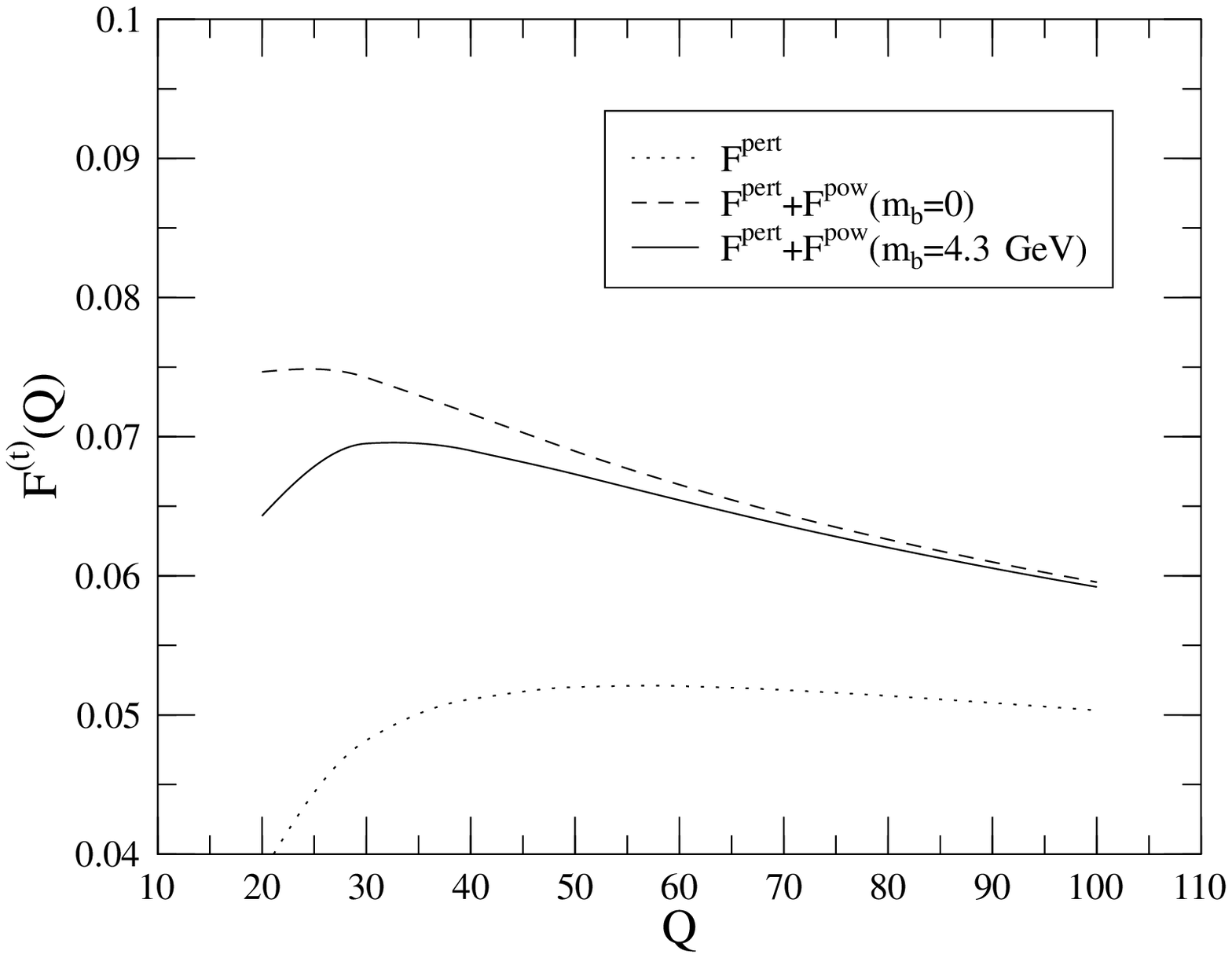} }
\end{minipage}
\caption{Mean value of (a) the $C$ parameter and (b) $1-T$, where $T$ is the
thrust. Dotted: next-to-leading order perturbative result, dashed:
perturbative + power correction without mass effect, solid: perturbative
+ power correction with mass effect.}
}

If we compare our results for the power corrections of tagged $b$ events to
the hadronization corrections obtained using Monte Carlo programs, we
find significantly smaller corrections from our model. For instance, the
double ratio $H_{b/l}^{(s)} = H^{(S)}_b/H^{(S)}_l$, where $H^{(S)}_x$ is the
hadron level value divided by the parton level value for event shape $S$
and for the quark flavour $x$ (heavy, or light), was determined in
Ref.~\cite{ALEPHbmass} for events in $Z^0$ decays using JETSET parton
shower model together with the Lund string fragmentation model.  In
this model the mass effects are introduced only by kinematic
constraints to the phase space at each parton branching in the shower
evolution.  Table~\ref{t:hadcorr} shows the values obtained for
$H_{b/l}^{(s)}$ using the two models.
\TABLE{
\label{t:hadcorr}
\renewcommand{\arraystretch}{1.2}
\begin{tabular} {|l||c|c|}
\hline
\hline
$S$   &   JETSET &  Disp.~model \\
\hline
\hline
$C$   &   1.175  &  1.000 \\
$t$   &   1.142  &  1.001 \\
\hline
\hline
\end{tabular}   
\caption{Comparison of hadronization corrections obtained using JETSET
and from the dispersive model via the double ratios $H_{b/l}^{(C)}$ and
$H_{b/l}^{(t)}$.}
}
The values for the double ratios obtained from
the dispersive approach are very close to 1, indicating that the
relative power corrections are very similar in the $b$-tagged and
$uds$-tagged samples.\footnote{This result depends very weakly on the
value of the Milan factor.} The situation is very different for the
JETSET estimates. The results, when translated for single ratios of
hadronization corrections on the tagged sample $H^{(S)}_b$, mean that
the hadronization correction from the dispersive approach is about half
of that from JETSET. It would be interesting to learn whether or not
this difference changes if the mass effects are taken into account in
the dynamics of the showering in JETSET, or it is rather due to the
different secondary decays of heavy hadrons. The latter case may
undermine the usefulness of estimating hadronization corrections from
power corrections for tagged $b$ events.

\section{Conclusion}

In this paper we have calculated the `naive contribution', as defined
in Ref.~\cite{DLMSMilan}, for the power corrections of $b$-tagged event
shapes thrust and $C$ parameter in \epem\ annihilation. At LEPI energy
the explicit analytic formulae predict about 7--10\,\% reduction of
the hadronization correction for the shapes obtained from $b$-tagged
samples with respect to the untagged case. This reduction effect will be
even more important for $t$-tagged samples at the NLC. For instance, for
$m_t = 175$\,GeV and $Q = 500$\,GeV, we expect the power correction to
be about two thirds of the corresponding massless case for the
$C$ parameter and $t$.

We also presented predictions for the mean values of the event shapes
by merging the perturbative and non-perturbative contributions.
The only non-perturbative input in the prediction is the
non-perturbative parameter $\alpha_0$, taken from fits of theoretical
predictions for the mean values of event shapes to those of light
primary quark samples. We found that the mass effect  in the power
correction for tagged $b$ samples is not too profound for centre-of-mass
energies above about 45\,GeV and is significantly smaller than the
hadronization correction predicted by the current version of JETSET,
which does not take into account the heavy-quark mass in the dynamics of
fragmentation, but includes decays of heavy hadrons.
The results may be used directly in heavy-quark mass measurements.

In our view the flavour independence of the strong coupling is
a sufficiently weak assumption so that if it is feasible with the
dispersive approach to calculate non-perturbative effects in hadronic
event shapes, then the primary quark mass effects should be
perturbatively calculable.  Therefore, our results, once confronted
with experiment, can serve as a further check of perturbatively
calculable hadronization corrections.

\acknowledgments
The author is grateful to G.~Salam and B.~Webber for useful
discussions, to C.~Oleari for helping with the {\tt zbb4} program and
to S. Catani for clarifying remarks on the manuscript.
This work was supported in part by the EU Fourth Framework Programme   
`Training and Mobility of Researchers', Network `Quantum      
Chromodynamics and the Deep Structure of Elementary Particles',      
contract FMRX-CT98-0194 (DG 12 - MIHT), as well as by the Hungarian   
Scientific Research Fund grant OTKA T-025482.

\def\etal{{\em et al.}}
\def\pl#1#2#3    {{\it Phys.\ Lett.\ }{\bf #1} (#2) #3~}
\def\eupc#1#2#3  {{\it Eur.\ Phys.\ J.\ }{\bf C #1} (#2) #3~}
\def\eupcdir#1#2#3  {{\it Eur.\ Phys.\ J. direct~}{\bf C #1} (#2) #3~}

\end{document}